# Analysis of thermally-induced effects in Planck Low Frequency Instrument

A. Mennella[1], M. Bersanelli[2], C. Burigana[3], D. Maino[4], R. Ferretti[5], G. Morgante[3,6], M. Prina[6], N. Mandolesi[3], C. Butler[3], L. Valenziano[3], F. Villa[3]

*On behalf of the LFI Consortium*

[1] *IFC-CNR, Milan, Italy*
[2] *Università di Milano, Dip. Di Fisica, Milano, Italy*
[3] *TESRE-CNR, Bologna, Italy*
[4] *Osservatorio Astronomico di Trieste, Trieste, Italy*
[5] *LABEN S.p.A., Vimodrone, Milan, Italy*
[6] *Jet Propulsion Laboratory, Pasadena, USA*

**Abstract.** The Planck mission will provide full-sky maps of the Cosmic Microwave Background with unprecedented angular resolution (~ 10') and sensitivity ($\Delta T$ / $T$ ~ $10^{-6}$). This requires cryogenically cooled, high sensitivity detectors as well as an extremely accurate control of systematic errors, which must be kept at μK level. In this work we focus on systematic effects arising from thermal instabilities in the Low Frequency Instrument. operating in the 30-100 GHz range. Our results show that it is of crucial importance to assure "in hardware" a high degree of stability. In addition, we provide an estimate of the level at which it is possible to reduce the contamination level in the observed maps by proper analysis of the Time Ordered Data.

## INTRODUCTION

Planck is an European Space Agency (ESA) mission to map spatial anisotropy in the Cosmic Microwave Background (CMB) over a wide range of frequencies with an unprecedented combination of sensitivity, angular resolution, and sky coverage. It consists of a Low Frequency Instrument (LFI) and a High Frequency Instrument (HFI) observing the sky through a common telescope [1, 2].

The LFI is constituted by an array of 54 radiometers actively cooled at 20 K that will collect the microwave radiation in four well defined frequency bands, centered at 30, 44, 70 and 100 GHz, while the HFI will image the sky in six frequency channels between 100 and 857 GHz with its array of 48 bolometric detectors cooled at 0.1 K. High thermal stability is required to avoid spurious systematic effects that need to be kept at μK level in order to retain the scientific value of the measured data.

To meet the cryogenic requirements a dedicated chain of cryo-coolers will be implemented on-board the Planck satellite. The 20 K stage, in common between the two instruments, will be provided by a Sorption Cooler [3], a vibration-less cryostat with a cooling capability of >1 W at 20 K. The heart of the cooler is composed by six hydride compressor beds in which hydrogen is alternatively absorbed and released as the temperature of the beds is modulated, thus creating a constant gas flow. The activity of such compressors generates second-order temperature oscillations that will propagate through the satellite and may affect the thermal boundaries between the spacecraft and the Planck instruments.

In this work we present the results of a preliminary assessment of the impact of temperature fluctuations of the 20 K stage of the LFI instrument. In our study we have estimated the maximum systematic error generated by such fluctuations of the LFI maps, taking into account the ability to reduce the level of these effect by applying destriping algorithms to the Time Ordered Data. Our results show that although the impact of such temperature instabilities can be reduced "in software" it is crucially important to guarantee "in hardware" a temperature stable at the mK level in the 20K stage.

## PLANCK THERMAL ENVIRONMENT

The instruments on board the Planck satellite will measure the CMB with a very high sensitivity, of the order of few μK per pixel at the end of a 14 month mission; in order to maintain the instrumental noise at very low levels a specifically designed chain of cryo-coolers is being currently developed. In Fig. 1 (a and b) we show a sketch representing the Planck thermal environment. The satellite, in particular, presents two main temperature stages (50 K and 300 K) thermally decoupled by three thermal shields ("V-grooves") at different temperatures which radiate heat into deep space. The instruments in the focal plane present different temperature stages ranging from 0.1 K (HFI bolometers) to 20 K (LFI radiometers).

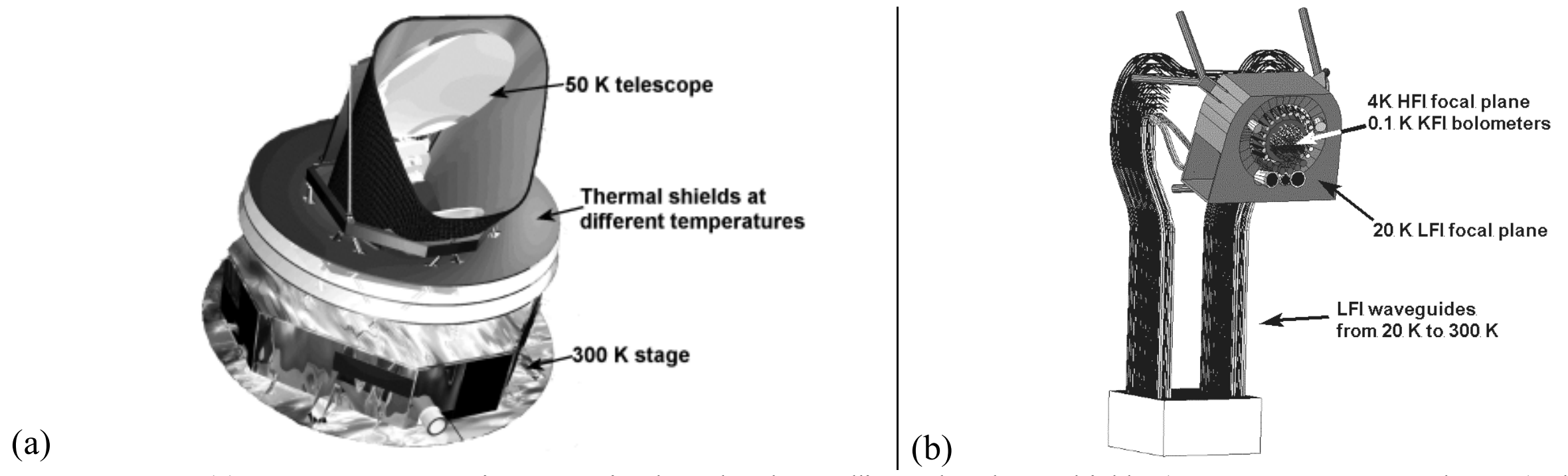

**FIGURE 1**. (a) Temperature environment in the Planck satellite. The three shields (at 140 K, 100 K and 50 K) thermally decouple the warm (300 K) and cold (50 K) satellite stages. (b) Temperatures of the Planck instruments.

Such a complex thermal environment clearly requires a very careful control of the temperature stability at the thermal boundaries between the spacecraft and the instruments, because any temperature variation will leave its signature in the measured data thus causing spurious systematic effects. The main source of temperature instability in Planck is represented by the Sorption Cooler (located in the 300 K Service Module) which drives the entire Planck cryo-chain and provides the 20 K environment to the LFI.

The temperature provided by the Sorption Cooler will not be perfectly stable, but will display oscillations caused mainly by pressure fluctuations in the compressor beds. These instabilities will impact on the behavior of many temperature-sensitive components of the LFI radiometers. As shown schematically in Fig. 2, each LFI radiometer is composed by a 20 K front-end stage, and a 300 K back-end. Each radiometer compares the signal from the sky to a stable 4 K reference load. In this work we consider the impact of temperature instabilities on the most sensitive 20 K front-end components, i.e. feed horns and orthomode transducers, the 4 K reference horn antenna and the front end RF amplifiers.

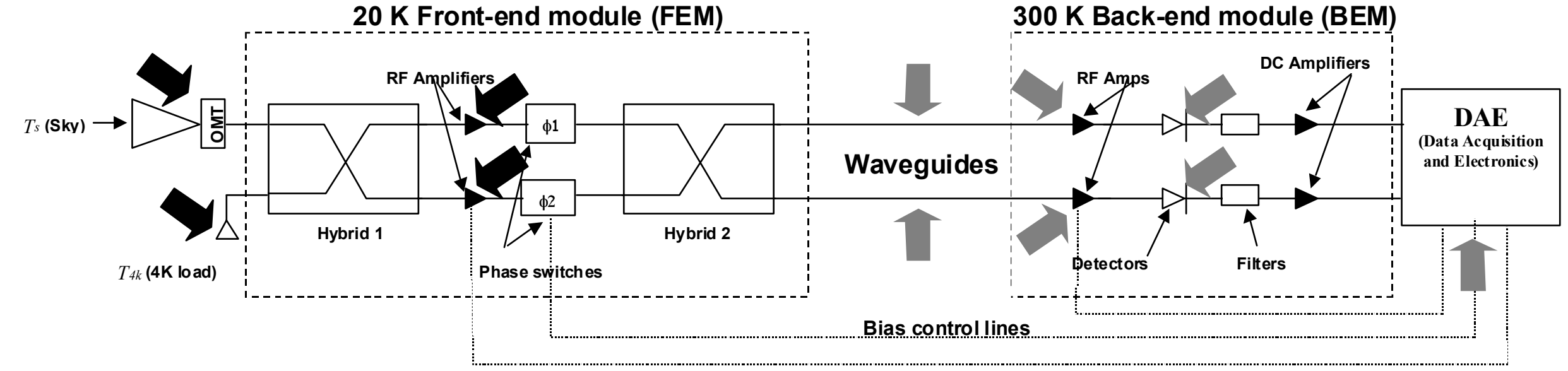

**FIGURE 2**. Schematic of Planck-LFI radiometers that highlights the components that are sensitive to temperature fluctuations. In this work we focus on the effect temperature instabilities on the most sensitive 20 K front-end components (black arrows).

Instability in the radiometer physical temperature will cause oscillations in the measured signal and, therefore, spurious anisotropies in the final maps. The impact of signal oscillations with a period that is not synchronous with the spacecraft spin (1 rpm) can be mitigated by applying so-called "destriping" algorithms to the Time Ordered Data, while spin-synchronous signal fluctuations will permanently affect the final maps. The high level of systematic error rejection required for Planck-LFI imposes strict limits on the maximum acceptable peak-to-peak error per pixel[1] caused by thermal effects ($\pm$ 0.8 $\mu$K for spin synchronous effects and $\pm$ 1.1 $\mu$K for other periodic effects).

# EFFECT OF FRONT END TEMPERATURE INSTABILITY

## Temperature fluctuations at the Sorption Cooler cold end (20 K)

Although experimental data concerning the shape of the physical temperature fluctuations caused by the Sorption Cooler are not available yet, recent simulations have provided a first insight into the effect of non-idealities in the compressor assembly behavior on the temperature stability of the 20 K cold end. In Figs. 3 and 4 we show two examples of the cold end temperature oscillation in two different scenarios: an ideally behaving cooler (Fig. 3) and a cooler with a non-homogeneous hydride distribution in one of the compressor beds (Fig. 4). The main result of these simulations is that Sorption Cooler-induced temperature fluctuations are characterized by a high number of harmonics of the compressor bed period (which currently set to 667 s). Furthermore a non ideal behavior in the compressor assembly results in an increased impact of higher and lower harmonics; in particular Fig. 4 shows a tail of higher harmonics in which it is visible (see Fig. 4b) a quasi-spin synchronous harmonic close to 60 s with an amplitude of the order of 1 mK.

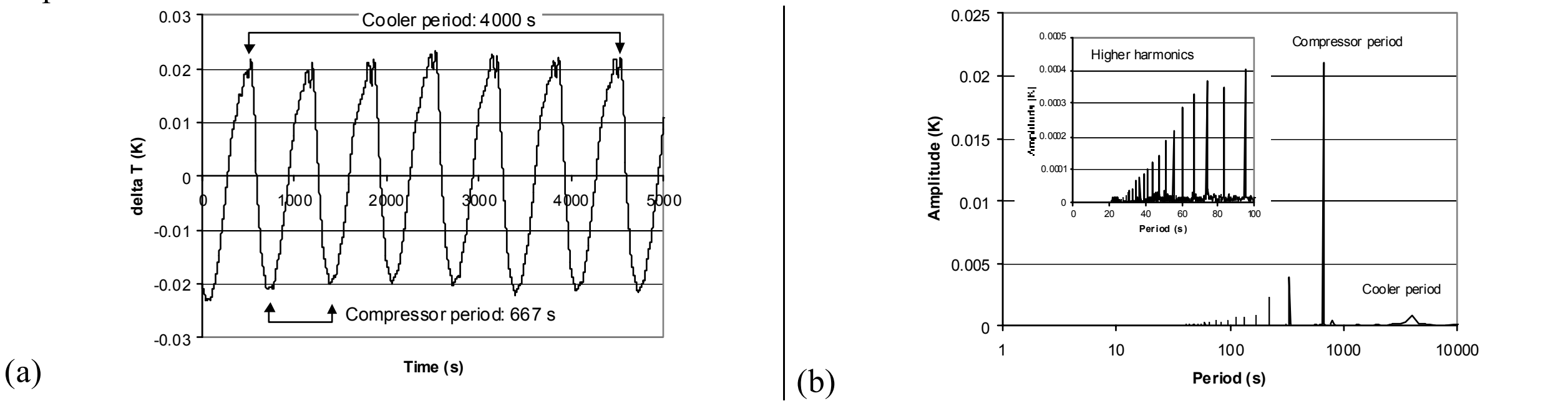

**FIGURE 3**. Simulation of temperature fluctuation at the sorption cooler cold end for an ideally-behaving set of compressor beds (6 equal compressors with homogeneous beds). (a) Oscillation in time domain. (b) Fourier transform. The inset is a zoom of the high frequency tail of the spectrum.

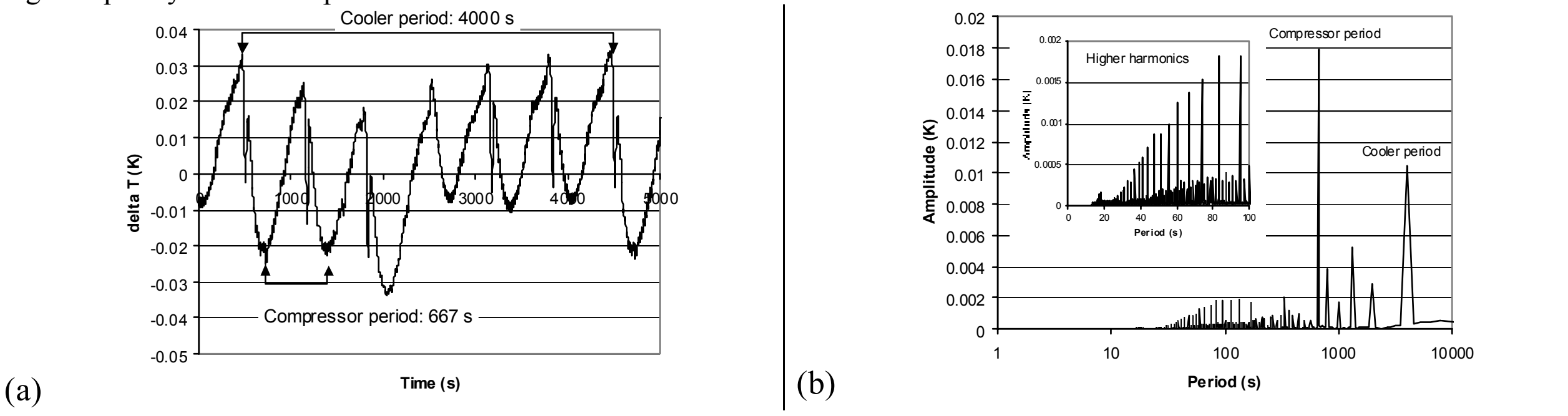

**FIGURE 4**. Same as in Fig. 3 for a Sorption Cooler characterized by 5 equal compressors with homogeneous beds and 1 compressor with non-homogeneous hydride distribution.

## Damping by LFI mechanical structure

In the previous section we have discussed the behavior of the physical temperature at the Sorption Cooler 20 K cold end. Now we want to evaluate how these fluctuations are transferred through the LFI mechanical structure to the LFI radiometers. The schematic in Fig. 5a shows the damping effect caused by the thermal mass and thermal resistance of the instrument, which acts as a low-pass filter for thermal oscillations. The net effect of this filter is that the actual temperature variation at the front-end radiometers display a lower peak-to-peak amplitude and a much more limited content of high frequency harmonics. In Fig. 5b we report results obtained using our current LFI thermal model which quantify the frequency dependence of the damping factor for the four LFI frequency channels.

---

[1] Here we refer to a pixel size equal to the physical beam size, $\theta_{FWHM}$. Note the LFI maps will be generated with a lower pixel size, of the order of $\theta_{FWHM}$ / 3, which corresponds to the data sampling resolution.

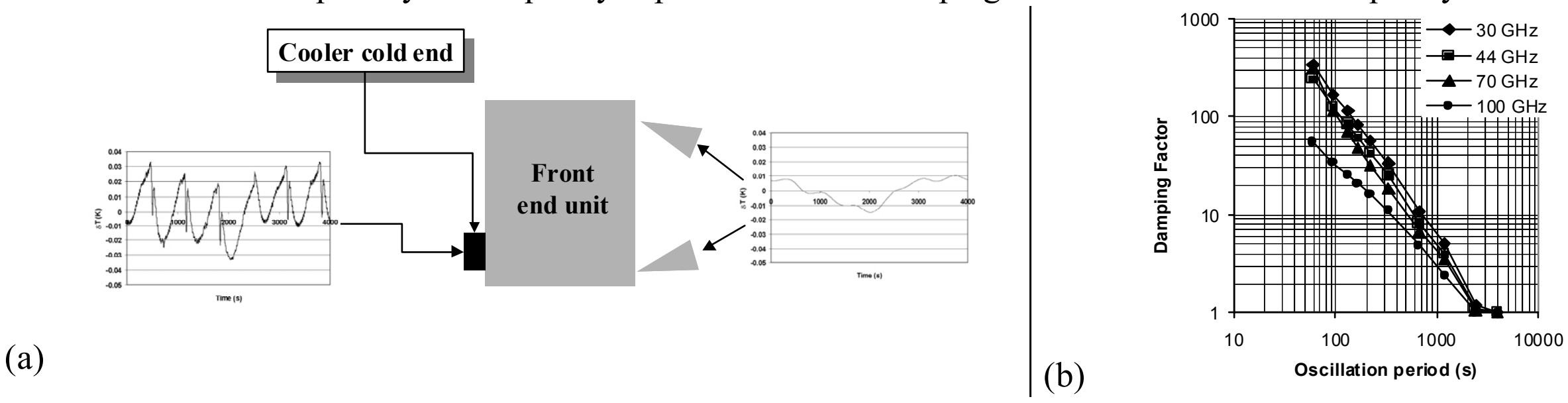

**FIGURE 5**. Damping of thermal by LFI mechanical structure. (a) The sketch shows that the LFI mechanical structure is a *low-pass* filter for thermal fluctuations. (b) Damping factor vs. fluctuation period for the four LFI frequency channels.

## Effect on LFI maps

The next step is to evaluate the effect of physical temperature fluctuations on the measured maps. In this paper we limit our analysis to the 30 GHz channel which appears also to be the most sensitive to temperature fluctuations.

Let us now review the steps involved in this process: (1) the instrument transfer function links a variation in the radiometer temperature to the oscillation in the measured signal; (2) the measurement redundancy during each scan yields a first damping of the signal oscillation; (3) the data stream of “averaged” sky circles is then “mapped” into the sky using the HEALPix hierarchical structure [4]; this yields a further damping of the peak-to-peak oscillation which is dependent on the map pixel-size; (4) maps from different receivers in the same frequency channel are co-added to obtain a single map for each frequency, with a further damping factor in the range 1 to 2.5 depending on the channel; (5) the peak-to-peak amplitude is rescaled on a pixel size equal to the beam width (of the order of 36' at 30 GHz). To perform this step we have first evaluated the r.m.s. amplitude of the effect from the map power spectrum (at the multipole $l$ corresponding to the beam angular resolution) and then we have calculated the peak-to-peak value assuming no change in the map distribution at the various angular scales.

Here we neglect the map co-adding step so that each map is relative to one particular LFI receiver, which has no impact on the results al lower frequencies (30 and 44 GHz).

In Fig. 6 we show 30 GHz maps (pixel size: 13.7') of the systematic error (in thermodynamic temperature) induced by front-end temperature fluctuations for the cases shown in Figs. 3 and 4. Map pixel coordinates have been produced using the LFI flight simulator, a code specifically developed to simulate the Planck mission. The figure shows that without applying destriping to the Time Ordered Data the maximum variation on the final maps is relevant, of the order of $\pm 16$ μK in the ideal case and about ten times greater in the more realistic case.

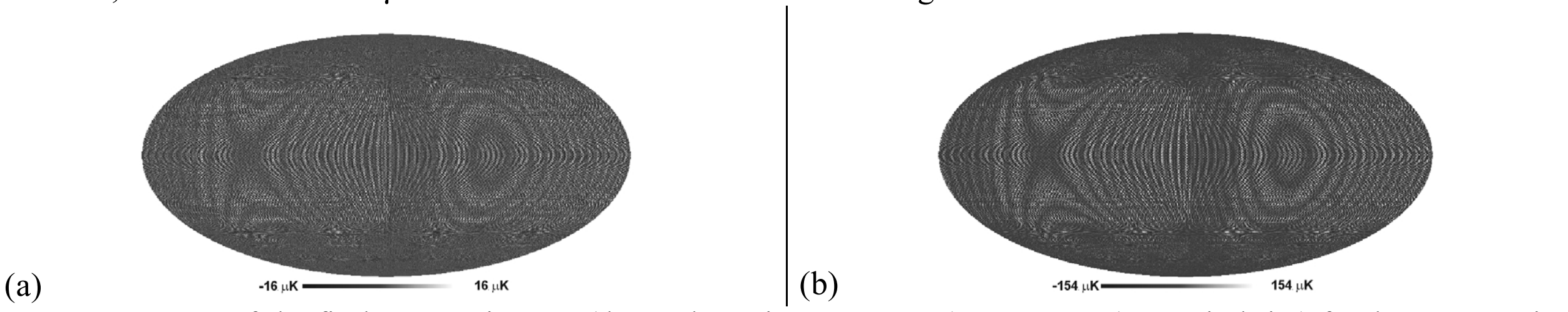

**FIGURE 6**. Map of the final systematic error (thermodynamic temperatures) at 30 GHz (13.7' pixel size) for the two Sorption Cooler scenarios considered in this note: (a) ideal Sorption Cooler, (b) Cooler with one non homogeneous bed.

Let us now evaluate the ability of reducing such systematic effects by applying a destriping algorithm to the Time Ordered Data[2]. The algorithm makes use of intersections in the sky scans (determined by the scanning strategy) to reduce the impact of periodic spurious signals (see [5] for further details). In general the destriping efficiency increases with the period of the spurious signal, while spin synchronous signals will not be altered by this procedure. This is shown clearly in Fig. 7, that shows the various damping factors (measurement redundancy, mapping and destriping) for signals having different periods.

[2] Note that, in general, the destriping code is applied to data streams that contain also the instrumental noise. Some tests have shown that the presence of a superimposed noise (white and white+1/f) has a negligible impact of the code ability to reduce the effect of periodic fluctuations.

In Table 1 we present a summary of the maximum peak-to-peak systematic error per pixel after destriping the maps shown in Fig. 6. The values summarized in the table indicate that even in the most favorable case the residual systematic is of the same order of the maximum error accepted from all thermal effects. Although this analysis is still preliminary and it is based on simulations of the Sorption Cooler behavior, it shows that a temperature stability at the 20 K cold end of order 10 mK would be needed to maintain the induced systematic errors at a negligible level.

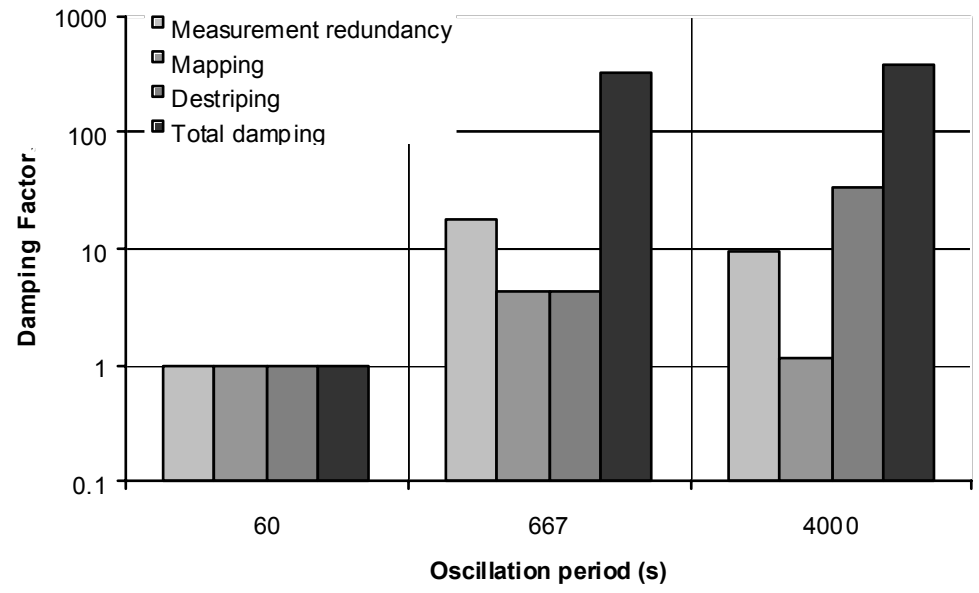

**FIGURE 7**. Typical damping factors for an (multiple pixel measurements and destriping) vs. fluctuation period. Spin synchronous fluctuations are not removed (damping = 1) while "longer" oscillations are reduced (after destriping) by a total factor of about 300.

**TABLE 1. Residual systematic error on final maps (on 36' pixel) caused by front-end temperature fluctuations**

| Shape of temperature fluctuation | **Peak-to-peak amplitude of systematic error in map (μK)** | |
|---|---|---|
| | Non spin synch. | Spin synch |
| 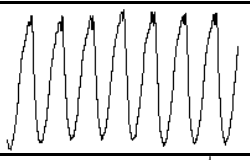 | 1.0 | 0.3 |
| 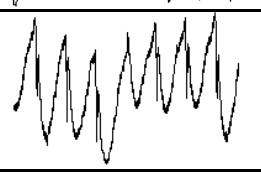 | 3.1 | 1.5 |

## CONCLUSIONS

In this paper we have presented a preliminary study of the impact of Sorption Cooler temperature fluctuations on the LFI measurements. Our results show that several factors contribute to reduce the impact of such fluctuations on the measured maps: the LFI mechanical structure that very efficiently reduces the amplitude of fast temperature oscillations (i.e. with a period less than 500s) propagating from the Sorption Cooler Cold end to the radiometers, the measurement redundancy during each scan circle and the application of destriping algorithms to the Time Ordered Data. Nevertheless, assuming non-homogeneous beds in the cooler, the expected maximum peak-to-peak systematic error per 36' pixel in the final maps at 30 GHz is at the 2-4 μK level: although apparently small, this is significantly more than the systematic error allocated to this effect. Future work will be focused at refining the analysis (including the use of experimental data from the Sorption Cooler breadboard, when available) and at identifying means to improve the temperature stability of the LFI front-end .

## ACKNOWLEDGMENTS

The HEALPix package use is acknowledged (see HEALPix home page at http://www.eso.org/science/healpix/). We also wish to thank the Planck LFI Data Processing Center for the support to the simulation work.

## REFERENCES

1. Mandolesi, N., *et al*, "Planck Low Frequency Instrument," in *Proceedings of 2K1BC Workshop on Experimental Cosmology @ mm-waves*, AIP Conference Proceedings, 2001 (this issue).
2. M. Bersanelli, N.Mandolesi, *Astrophisical Letter and Communication*. **37**, 171-180 (2000).
3. Bhandari, P., *et al*, *Astrophysics Letters and Communications* **37**, 227-237 (2000).
4. Gorski, K.M., Hivon, E., Wandelt, B.D., "Analysis Issues for Large CMB Data Sets", *Proceedings of the MPA/ESO Conference on Evolution of Large-Scale Structure: from Recombination to Garching*," edited by Banday, A.J., Sheth, R.K., Da Costa, L., 37-42, 1998.
5. Maino, D., *et al*, Astronomy and Astrophysics Supplement Series **140**, 383-391 (1999).